\documentclass{article}

\usepackage{arxiv}

\usepackage[utf8]{inputenc} 
\usepackage[T1]{fontenc}    
\usepackage{hyperref}       
\usepackage{url}            
\usepackage{booktabs}       
\usepackage{amsfonts}       
\usepackage{nicefrac}       
\usepackage{microtype}      
\usepackage{lipsum}
\usepackage{graphicx}
\usepackage{float}
\usepackage{amsmath}
\graphicspath{ {./images/} }

\title{Patch based Transformation for Minimum Variance Beamformer Image Approximation Using Delay and Sum Pipeline}

\author{
Sairoop Bodepudi \\
Indian Institute of Technology Palakkad \\
Kerala, India \\
\texttt{sairoopb@gmail.com}\\
\And
A. N. Madhavanunni \\
Indian Institute of Technology Palakkad \\
Kerala, India\\
\texttt{121813001@smail.iitpkd.ac.in}\\
\And
Mahesh Raveendranatha Panicker\\
Indian Institute of Technology Palakkad \\
Kerala, India \\
\texttt{mahesh@iitpkd.ac.in}
}

\date{}
\begin{document}
\maketitle
\begin{abstract}
In the recent past, there have been several efforts in accelerating computationally heavy beamforming algorithms such as minimum variance distortionless response (MVDR) beamforming to achieve real-time performance comparable to the popular delay and sum (DAS) beamforming. This has been achieved using a variety of neural network architectures ranging from fully connected neural networks (FCNNs), convolutional neural networks (CNNs) and general adversarial networks (GANs). However most of these approaches are working with optimizations considering image level losses and hence require a significant amount of dataset to ensure that the process of beamforming is learned. In this work, a patch level U-Net based neural network is proposed, where the delay compensated radio frequency (RF) patch for a fixed region in space (e.g. 32x32) is transformed through a U-Net architecture and multiplied with DAS apodization weights and optimized for similarity with MVDR image of the patch. Instead of framing the beamforming problem as a regression problem to estimate the apodization weights, the proposed approach treats the non-linear transformation of the RF data space that can account for the data driven weight adaptation done by the  MVDR approach in the parameters of the network. In this way, it is also observed that by restricting the input to a patch the model will learn the beamforming pipeline as an image non-linear transformation problem.
\end{abstract}

\section{Introduction}
Medical ultrasound (US) is a diagnostic tool that finds applications in evaluating almost the entire human body for example, cardiac, abdominal, fetal, musculoskeletal, breast imaging, etc. Advantages of US over other imaging modalities include real-time imaging capabilities, can be done bedside for sick patients, cost-effectiveness, and absence of harmful radiation \cite{szabo}. Ultrasound Imaging involves computationally intensive signal processing, where a piezo-electric transducer array is employed to transmit and receive pulsed waveforms, which are then beamformed to produce an output image \cite{kai_bfevolution}. There have been many beamforming algorithms developed over the last decade to improve image quality. The most widely employed beamforming algorithm has been the data independent delay and sum (DAS) beamforming with dynamic receive focusing where the delay compensated signals are weighted according to the location of the pixels with respect to transducer geometry and summed to form the final beamformed signal \cite{szabo,kai_bfevolution}. The DAS has been popular due to low complexity and realtime image reconstruction even though the image quality has not been the best \cite{garcia_das}.\\

However, the images reconstructed using DAS algorithm suffer from poor resolution and contrast. To overcome the inherent issues with DAS, data adaptive beamforming schemes based on Capon/Minimum variance distortionless response (MVDR) techniques \cite{mvdr} have also been proposed. Unlike DAS, the apodization weights are estimated from the received data for MVDR beamforming. The schemes provide better resolution and improved clutter rejection, but the high computational complexity restricts the real-time implementation. A non-linear beamforming scheme called filtered delay multiply and sum (F-DMAS) has also been proposed \cite{dmas}, where the delay compensated channel data are cross multiplied in all possible pair-wise combinations (excluding the self-products) and then summed to form the final image. The beamformed signal is filtered to extract the second harmonic component that is generated from multiplication which results in better lateral resolution and contrast \cite{dmas}. However, irrespective of all the developments, DAS has been still the defacto beamforming algorithms in all the commercial diagnostic ultrasound imaging system.\\

With the advent of deep learning (neural networks) and its power of being a powerful function approximator leveraging its ability to learn a hierarchy of nonlinear parametrized operations, the potential of the same in the field of ultrasound imaging is also being investigated. In the recent past, neural networks have been employed to replace or aid the conventional signal processing pipeline \cite{eldar_dlus}. A few examples of this are, in \cite{dahl_dlspecklered} where neural networks (NNs) are used in beamforming to minimize speckle noise, in \cite{luitjen_dl} a learning-based model is used to make an adaptive beamformer where the apodization weights of pixel input data from the transducer are controlled by a FCNN. Deep neural networks have been used to reconstruct high quality brightness-mode (B-mode) ultrasound images from sub-sampled RF data \cite{khan_dlusbsample}. The basic pipeline however in this model is similar to a typical DAS signal processing pipeline. \\

There have also been other works such as \cite{khan_dlcompbeam} where the authors use a deep learning model for end to end transformation from raw RF channel data to B-Mode output image along with the incorporation of various other beamforming techniques like the MVDR Beamformer, imposing deconvolution beamformer and finally also applying causality condition. The authors also incorporate compressive beamforming by using subsampled RF data to generate the output in phase - quadrature (IQ). There have also been works like \cite{eldar_dlsubsampbeam} where the authors try to resolve the problem of requiring high sampling rates for high quality image construction and use deep learning to produce the B-Mode Image rather than fully sampled RF data from the sub-sampled data. In \cite{garcia_iqdlbeam}, complex convolutional networks are used for reconstruction of images from IQ data as they are low cost strategies and can be used for ultra fast ultrasound imaging. This is the first time, where neural network accelerators for IQ data has been proposed. There have also been approaches of beamforming for the creation of multiple images by just varying some hyper parameters as in \cite{khan_switchablebeam} where rather than having multiple models for reconstruction similar to various algorithms they have a single model which would generate the corresponding image.\\

In addition to replace beamforming by using neural network accelerators, there have also been a lot work in improving the image quality of the constructed image using Deep Learning algorithms such as \cite{sim_cycleGAN} where the authors used a variation of the cycle consistency generative adversarial network (cycleGAN) \cite{zhu_cycleCAN} along with incorporation of the patch based discriminator and Optimal Transport. The authors show that by doing this the quality of generated Images can be improved considerably. Other works such as \cite{khan_UnpervisedUS} where the authors used cycleGAN to model the super resolution in ultrasound imaging by artifact removal in an unsupervised fashion. The authors \cite{khan_UnpervisedUS} also mention that this work is able to tackle other similar objective works involving deconvolution, speckle removal amongst many others. There have also been self-supervised learning based approaches for super resolution  keeping in mind the lack of available high resolution images in \cite{liu_perceptionCycleGAN} and thus prove to be superior to state of the art super resolution algorithms for Ultrasound image super resolution on some datasets.

\begin{figure*}[!t]
\centerline{\includegraphics[width=0.82\paperwidth]{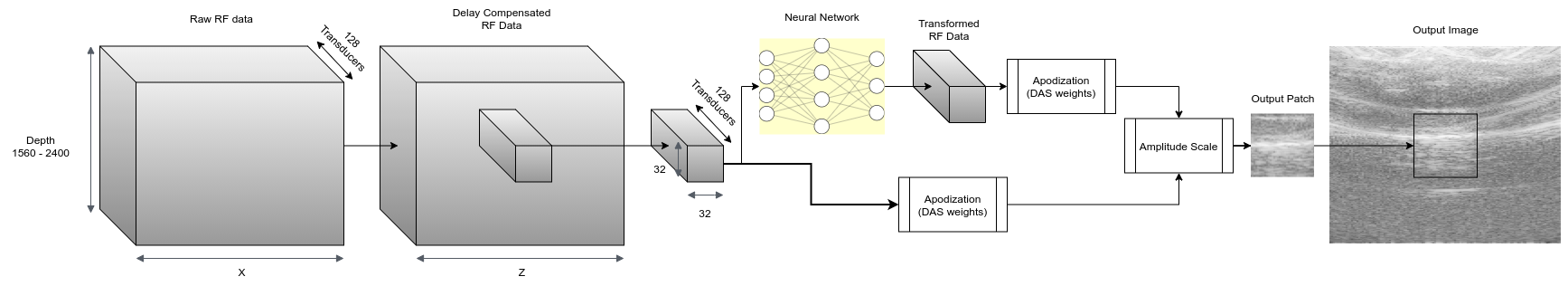}}
\caption{Illustration of the entire pipeline}
\label{fig1}
\end{figure*}

\section{Motivation and Contribution}
It is evident that there have been a plethora of efforts in employing NNs for accelerating and improving the beamforming process employing the non-linear functional approximation ability of NNs. However most of these approaches are working with optimizations considering image level losses and hence require a significant amount of dataset to ensure that the process of beamforming is learned. Also, most of the efforts have been to estimate the apodization weights by formulating the NN as a regression problem. In the proposed approach we use a patch based transformation from the delay compensated data to account for the data driven weight aspect of the MVDR beamforming pipeline. This is to make sure that after a transformation from the RF data, a simple and fast DAS beamforming should produce high quality images comparable to that of MVDR produced images. At the final stage we scale the output image based on the output of DAS since DAS output images can find the high level structure and the transformation helps in describing the details. \\

We also train the model from different types and locations of body parts so that the network does not learn some sort of mapping which is effective to a specific part and performs well on different locations. Most of the preexisting models are trained on a specific location of the body, but the proposed approach is more general. Rather than learning the characteristics of a specific part or locations and their relation with the input RF data, the proposed model learns the more important transformation pipeline that the MVDR  captures in the weight adaptation. Moreover in comparison with other methods our method can be parallelized such that the DAS is to be applied only over a patch instead of the entire image.  

\begin{figure*}[!t]
\centerline{\includegraphics[width=0.6\paperwidth]{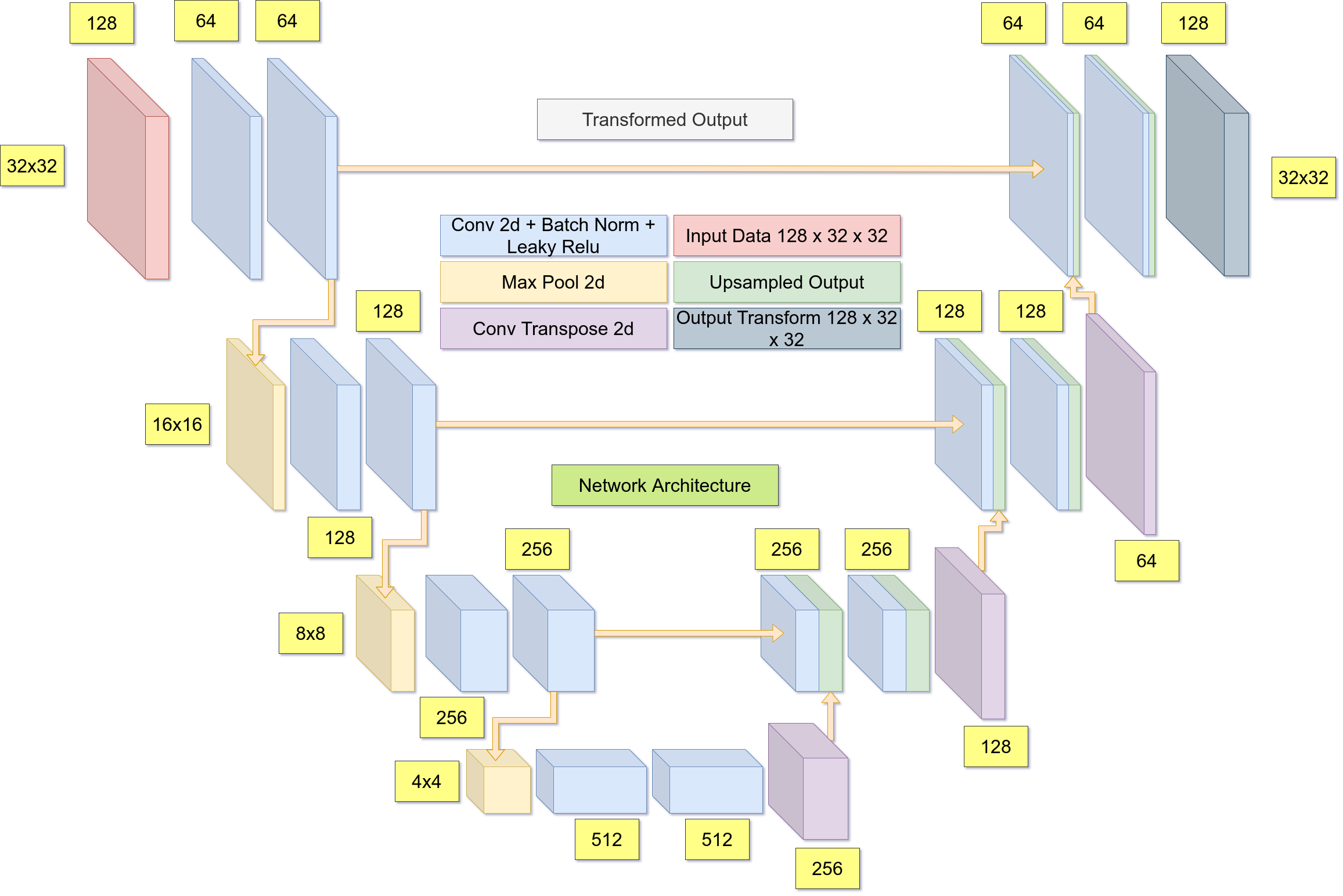}}
\caption{Illustration of the entire network architecture for a 32 x 32 patch size input}
\label{fig2}
\end{figure*}

\section{Proposed Approach}
\subsection{Overview}
Inspired by the success of U-Net \cite{u-net} especially in tasks relating to Medical Image Segmentation the underlying backbone for the delay compensated RF channel data transformation is similar to the U-Net model. Since Neural Networks have influenced the previous decade being powerful domain transformers the objective of the network here is to model the weight adaptation that the Delay and Sum algorithm fails to account and that the Minimum Variance beamformer does, in an efficient and computationally feasible manner.\\

Towards this end, we pass the delay compensated RF channel data to the network which after performing the appropriate transformation is passed onto the DAS beamformer. The motivation for the same apart from the previously mentioned reasons is to account for the system configurations and at the same time some geometric properties of the sensors which might not have been captured by the network. Thereby our model is not bottle-necked by the weight adaptation of the data-driven beamformers and can perform on par in a more efficient manner.

\subsection{Neural Network Architecture}
In the proposed model, we first apply the convolution operations followed by a non linearity function accompanied by pooling layers. At every point, the input's height and width are reduced by half respectively but the number of channels are doubled till a prefixed threshold. This representation can be understood as the latent space representation of the RF channel data from which it is upsampled and concatenated with the partially transformed input of the same dimensions. This is to make sure that not a lot of the input is lost and better propagation of the gradients.

This transformed RF data is sent to the DAS apodization and is scaled according to the DAS beamformed image to obtain the final output. For input RF channel data $z$ and network $\Phi_{\theta}$ parametrized over $\theta$, the pipeline for a patch of the image can be mathematically written as
\begin{equation}
    z' = \Phi_\theta (z)
\end{equation}
where $z'$ is the transformed RF channel data which is then sent into the DAS beamformer $B$
\begin{equation}
    h = B (z')
\end{equation}
where $h$ is the output image which is to be rescaled with the $Scale$ function based on the Delay and Sum beamformer output as
\begin{equation}
    y' = Scale(h, B(z))
\end{equation}
where $y'$ is the final predicted output image and the $Scale$ function takes the input image to be scaled and the reference image for the scaling range.

The network is trained by solving the following optimization problem
\begin{equation}
    min_\theta\sum_{i=1}^N\alpha*||y'_i - y_i||_1 - \beta *SSIM(y'_i, y_i)
\end{equation}
where $y'_i=Scale(B(\Phi_\theta(z_i)), B(z_i))$ for the datapoint $\{z_i, y_i\}$ and $N$ is the total number of datapoints in the dataset. Here the loss function is defined as a hybrid consisting of Structural Similarity Index Loss (SSIM) \cite{ssim} and Mean Absolute Error (MAE).

We used the structural similarity loss so that the network can learn the structure if present within the patches and also the MAE, so that the image formed is closer to the original in the lower-level details. This was done by imposing a weighting between the loss functions depending on the tradeoff between the lower-level details and the higher-level structure.

The tradeoff was controlled in the model training by changing the hyperparameters $\alpha$ and $\beta$ where having a higher $\alpha$ results in more detail at the lower level whereas a higher $\beta$ would aim for higher structure at the lower level.

\subsection{Training details}
The model was trained in multiple iterations, the first was a basic pipeline which was trained on a subset of the data consisting of 32 images to check the efficacy of the model and performance in the low shot setting. The final iteration was trained on the entire training dataset for 14000 steps for a batch size of 64 where the total number of images that were used were 84 including the validation dataset.

The optimizer used for the training of the Neural network is the Adam optimizer \cite{adam} with the default hyperparameter settings since it has low memory requirements and is adaptive and is suitable for problems that have large variance in the gradients which we expect in this problems due the large range of values for both the input and the output.

The model was trained using a weighted loss function of SSIM and MAE with a weighting of 0.9 to MAE and 0.1 for the SSIM. The reason for this was that at the patch level it is more important for the network to learn the lower details rather than the structure.

The training data was sampled using random sampler to increase the generalization on the patches rather than the model learning to overfit the data with a training validation split of 0.8. The model is thus producing good results on this training/validation dataset and it is extending to the test set as well in visual testing.

\section{Results}
We see that by constraining the model to learn to construct images from just 84 samples when we choose the patch size of 32x32 for the patch model whose output is then stitched together to form the complete output B-Mode image, we are able to learn in a few shot setting and produce good results and also ensure that the model is not exposed to the entire image which would make the model learn a direct mapping and not the beamforming pipeline.

\begin{figure}[H]
    \centering
    \includegraphics[scale=0.5]{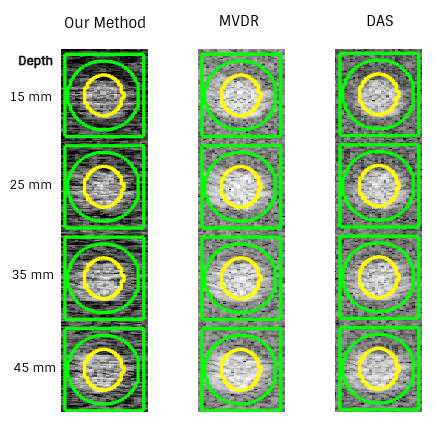}
    \caption{Beamformed images for a dataset with 4 cysts at different depths using our proposed method alongside the MVDR and DAS beamformers}
    \label{fig:results}
\end{figure}

We visualise the results for a single plane wave for 4 cysts in \ref{fig:results} and in \ref{tab:results} we have quantitative values for the contrast ratio metrics of different methods (i.e MVDR Beamformer, DAS Beamformer) in comparison to our method at different depths. We use the contrast ratio metrics defined in \cite{cubdl} as defined below where $\mu_{1}$ is the mean of the inner circle containing the cyst and $\mu_{2}$ is the mean of the outer circle containing the area surrounding the cyst as well.
\begin{equation*}
\text { Contrast Ratio }=20 \log _{10}\left(\frac{\mu_{1}}{\mu_{2}}\right)
\end{equation*}

\begin{table}[H]
	\caption{Contrast ratio for the proposed method at 4 different depths}
	\centering
	\begin{tabular}{llll}
		\toprule
		Depth (mm)     & Our Method     & MVDR  & DAS \\
		\midrule
		15     & +0.667  & +0.514   & +0.550   \\
		25     & +0.604  & +0.460   & +0.498   \\
		35     & +0.577  & +0.483   & +0.519   \\
		45     & +0.588  & +0.456   & +0.497   \\
		\bottomrule
	\end{tabular}
	\label{tab:results}
\end{table}

\section{Conclusions and Future Work}
In this work, we propose a neural network model for ultrasound image reconstruction on par with that of an MVDR beamformer at a higher speed. Throughout this work, we have seen that with a patch-based approach for solving the beamforming pipeline, the performance and the speed are providing results on par with state-of-the-art models. With the patch size being smaller, the DAS also takes lesser time in comparison to the processing of a larger image and therefore an overall reduction of image formation time which was achieved by using a hybrid loss of SSIM and mean absolute error at the patch level. 

Future work would include experimenting on more Neural network architectures. Also, more studies can be made relating to the optimal patch size which can play a huge role in determining the tradeoff between lower-level details and the structure along with determining how well the network can interpret and generalize the beamforming pipeline.

\bibliographystyle{unsrt}  

\end{document}